\documentclass[pdflatex,sn-mathphys-num]{sn-jnl}

\usepackage{graphicx}
\usepackage{multirow}
\usepackage{amsmath,amssymb,amsfonts}
\usepackage{amsthm}
\usepackage{mathrsfs}
\usepackage[title]{appendix}
\usepackage{xcolor}
\usepackage{textcomp}
\usepackage{manyfoot}
\usepackage{booktabs}
\usepackage{algorithm}
\usepackage{algorithmicx}
\usepackage{algpseudocode}
\usepackage{listings}

\usepackage{braket}
\usepackage{amsfonts}
\usepackage{bm}

\theoremstyle{thmstyleone}

\newtheorem{proposition}{Proposition}

\newtheorem{corollary}{Corollary}
\newtheorem*{example*}{Example}
\newtheorem*{remark*}{Remark}
\newtheorem*{remarks*}{Remarks}

\newcommand{\dd}{\,\mathrm{d}}
\newcommand{\tr}{\,\mathrm{tr}}
\newcommand{\ee}{\,\mathrm{e}}
\newcommand{\E}{\,\mathbb{E}}

\begin{document}

\title[Relative entropy]{Average relative entropy of random states}
\author{\fnm{Lu} \sur{Wei}\vspace{-0.4cm}}

\affil{\orgname{\center Department of Computer Science \\ Texas Tech University} \\ \orgaddress{\city{Lubbock}, \state{Texas} \postcode{79409}, \country{USA}}}

\abstract{Relative entropy serves as a cornerstone concept in quantum information theory. In this work, we study relative entropy of random states from major generic state models of Hilbert-Schmidt and Bures-Hall ensembles. In particular, we derive exact yet explicit formulas of average relative entropy of two independent states of arbitrary dimensions from the same ensemble as well as from two different ensembles. One ingredient in obtaining the results is the observed factorization of ensemble averages after evaluating the required unitary integral. The derived exact formula in the case of Hilbert-Schmidt ensemble complements the work by Kudler-Flam~(2021 Phys Rev Lett {\bf 126} 171603), where the corresponding asymptotic formula for states of equal dimensions was obtained based on the replica method.}


\maketitle

\section{Introduction and Main Results}\label{sec:intro}
Random states find diverse applications in modern quantum science including quantum circuit complexity~\cite{Brandao21}, quantum device benchmark~\cite{Choi23}, and entanglement estimation~\cite{Page93}. Different assumptions on the structure of Hilbert space lead to different random state models with the most prominent ones being the Hilbert-Schmidt ensemble~\cite{Page93,HW25}, the Bures-Hall ensemble~\cite{Kumar19,WHW25}, and the fermionic Gaussian ensemble~\cite{Bianchi21,HW23}. Exact statistical characterization of entropic metrics, such as von Neumann entropy and entanglement capacity, over these ensembles has been intensively studied for a single random state from one ensemble~\cite{Bianchi21,Foong94,HWC21,HW23,HW25,Page93,Ruiz95,Kumar19,VPO16,Wei17,Wei20,Wei20BHA,Wei20BH,Wei23,WHW25}.
Much less explored are distinguishability metrics of two random states from the same or two distinct ensembles. In such scenarios, the most fundamental metric is the relative entropy~\cite{Kudler21,Leditzky16}.

A density matrix $\rho$ is a positive semidefinite matrix in a Hilbert space satisfying
\begin{equation}\label{eq:dm}
\tr\rho=1.
\end{equation}
The relative entropy between two density matrices $\rho$ and $\sigma$ of the same dimension is defined as~\cite{Kudler21,Leditzky16}
\begin{equation}\label{eq:red}
D(\rho||\sigma)=\tr\left(\rho\left(\ln\rho-\ln\sigma\right)\right).
\end{equation}
In classical information theory, relative entropy is known as Kullback-Leibler divergence~\cite{Chen25} that measures the difference between two probability density functions. As a distinguishability metric of density matrices of quantum states, the relative entropy~(\ref{eq:red}) satisfies various information-theoretic properties~\cite{Leditzky16}. These include nonnegativity
\begin{equation}
D(\rho||\sigma)\geq0
\end{equation}
with equality $D(\rho||\sigma)=0$ for identical density matrices $\rho=\sigma$, and monotonicity
\begin{equation}
D(\rho||\sigma)\geq D\left(\Psi(\rho)||\Psi(\sigma)\right)
\end{equation}
with $\Psi$ denoting any completely positive trace-preserving map.

Clearly, the relative entropy~(\ref{eq:red}) becomes a random variable when considering random states. A natural first task is to understand the typical behavior of relative entropy by computing its average value. As discussed in~\cite{Kudler21}, closed-form formulas of relative entropy of random states are useful in several areas of quantum information theory including quantum hypothesis testing~\cite{Hiai91} and eigenstate thermalization hypothesis~\cite{He17}. Despite the importance, existing results in this direction are rather limited. In~\cite{Kudler21}, using some large-dimensional estimation and the replica trick, an approximate expression of average relative entropy of Hilbert-Schmidt ensemble~(\ref{eq:HS}) was obtained. In this work, we derive exact yet explicit formulas of average relative entropy of the Hilbert-Schmidt ensemble~(\ref{eq:HS}) and another generic state model of Bures-Hall ensemble~(\ref{eq:BH}). The more interesting case of relative entropy between the two ensembles is also considered, where we derive the corresponding exact average formulas. 

From the definition~(\ref{eq:red}), the average relative entropy of random states is given by
\begin{equation}\label{eq:rem}
\E\!\left[D(\rho||\sigma)\right]=\E\!\left[\tr(\rho\ln\rho)\right]-\E\!\left[\tr(\rho\ln\sigma)\right].
\end{equation}
The first term $\E\!\left[\tr(\rho\ln\rho)\right]$ is, up to a negative sign, the average entanglement entropy of reduced density matrix $\rho$ of a bipartite system. In the literature, an exact formula of average entanglement entropy for the Hilbert-Schmidt (HS) ensemble~(\ref{eq:HS}) is obtained in~\cite{Page93,Foong94,Ruiz95,HW25} as
\begin{equation}\label{eq:HSm}
\E\!\left[-\tr(\rho_{\rm{HS}}\ln\rho_{\rm{HS}})\right]=\psi_{0}(mn+1)-\psi_{0}(n)-\frac{m+1}{2n}
\end{equation}
and that for the Bures-Hall (BH) ensemble~(\ref{eq:BH}) is obtained in~\cite{Kumar19,Wei20BHA} as
\begin{equation}\label{eq:BHm}
\E\!\left[-\tr(\rho_{\rm{BH}}\ln\rho_{\rm{BH}})\right]=\psi_{0}\!\left(mn-\frac{m^2}{2}+1\right)-\psi_{0}\!\left(n+\frac{1}{2}\right),
\end{equation}
where the dimension of density matrices $\rho_{\rm{HS}}$ and $\rho_{\rm{BH}}$ is $m$ with parameter $n$, cf.~(\ref{eq:a}). The function $\psi_{0}$ denotes the digamma function~\cite{Brychkov}
\begin{equation}\label{eq:dg}
\psi_{0}(x)=\frac{\dd}{\dd x}\ln\Gamma(x),
\end{equation}
where, for a positive integer $l$, one has
\begin{subequations}\label{eq:dgs}
\begin{eqnarray}
&&\psi_{0}(l)=-\gamma+\sum_{k=1}^{l-1}\frac{1}{k} \\
&&\psi_{0}\!\left(l+\frac{1}{2}\right)=-\gamma-2\ln2+2\sum_{k=0}^{l-1}\frac{1}{2k+1}
\end{eqnarray}
\end{subequations}
with $\gamma\approx0.5772$ being the Euler's constant. The result~(\ref{eq:HSm}) is the celebrated Page's mean entropy formula~\cite{Page93}.

The computation of average relative entropy now boils down to computing
\begin{equation}\label{eq:rem2}
\E\!\left[\tr(\rho\ln\sigma)\right]
\end{equation}
in~(\ref{eq:rem}) that involves two independent random density matrices. The expected value~(\ref{eq:rem2}) quantifies the average lack of information of $\rho$ under $\sigma$, where $\sigma$ is typically a simpler model associated with a more general model $\rho$. For example, if $\rho$ is some full probabilistic model, then $\sigma$ can be a reduced-order approximate model, which is less informative but is often useful to accelerate computations~\cite{Chen25}. Consequently, in addition to the two cases $\E\!\left[\tr(\rho_{\rm{HS}}\ln\sigma_{\rm{HS}})\right]$ and $\E\!\left[\tr(\rho_{\rm{BH}}\ln\sigma_{\rm{BH}})\right]$, we also focus on the case $\E\!\left[\tr(\rho_{\rm{BH}}\ln\sigma_{\rm{HS}})\right]$ instead of the option $\E\!\left[\tr(\rho_{\rm{HS}}\ln\rho_{\rm{BS}})\right]$. This is because the Bures-Hall ensemble is a natural generalization of the Hilbert-Schmidt ensemble as can be seen by comparing their density matrix constructions~(\ref{eq:maHS}) and~(\ref{eq:maBH}). Exact formulas of~(\ref{eq:rem2}) in the considered three cases are the main results of this work. The results are summarized in the three propositions below, where the proofs are found in Section~\ref{sec:proof}.

\begin{proposition}\label{prop:HS}
For density matrices $\rho_{\rm{HS}}$ and $\sigma_{\rm{HS}}$ of Hilbert-Schmidt ensemble~(\ref{eq:HS}) of dimension $m$ with parameters $n_1=m+\alpha_1$ and $n_2=m+\alpha_2$, respectively, the average relative entropy~(\ref{eq:rem}) is given by
\begin{eqnarray}\label{eq:r1}
\E\!\left[D(\rho_{\rm{HS}}||\sigma_{\rm{HS}})\right]&=&\psi_{0}\!\left(mn_{2}\right)-\psi_{0}\!\left(mn_{1}+1\right)+\psi_{0}\!\left(n_{1}\right)-\frac{1}{m}\left(n_{2}\psi_{0}\!\left(n_{2}\right)-\alpha_{2}\psi_{0}\!\left(\alpha_{2}\right)\right) \nonumber \\
&&+~\!\frac{m+2n_{1}+1}{2n_{1}}.
\end{eqnarray}
\end{proposition}

It is worth noting a limiting formula of Proposition~\ref{prop:HS}; the result is presented as a corollary below.
\begin{corollary}\label{coro:HS}
For density matrices $\rho_{\rm{HS}}$ and $\sigma_{\rm{HS}}$ of dimension $m$ with parameters $n_{1}$ and $n_{2}$, respectively, in the regime
\begin{equation}\label{eq:lim}
m\to\infty,~~~~n_{i}\to\infty,~~~~\frac{n_{i}}{m}=c_{i}\geq1,~~~~i=1,2
\end{equation}
with $c_{1}$ and $c_{2}$ being fixed constants, the limiting formula of average relative entropy~(\ref{eq:r1}) is
\begin{equation}\label{eq:HSlim}
\E\!\left[D(\rho_{\rm{HS}}||\sigma_{\rm{HS}})\right]=\left(c_{2}-1\right)\ln\left(1-\frac{1}{c_{2}}\right)+\frac{1}{2c_{1}}+1+\mathcal{O}\left(\frac{1}{m}\right).
\end{equation}
\end{corollary}
Corollary~\ref{coro:HS} is readily obtained from Proposition~\ref{prop:HS} by applying, in the limit~(\ref{eq:lim}), the asymptotic expansion of digamma function~\cite{Brychkov}
\begin{equation}\label{eq:dga}
\psi_{0}(x)\sim\ln x-\frac{1}{2x}-\sum_{k=1}^{\infty}\frac{B_{2k}}{2k~\!x^{2k}},~~~~~~x\to\infty
\end{equation}
with $B_{2k}$ being Bernoulli numbers.

The limiting formula~(\ref{eq:HSlim}) in the special case $c_{1}=c_{2}$, i.e., states of equal parameters $n_{1}=n_{2}$, has appeared recently as the main finding in~\cite{Kudler21}. Despite being derived using the replica trick along with some approximations, the result~\cite[Eq.~(23)]{Kudler21} precisely captures the asymptotic behavior of average relative entropy of Hilbert-Schmidt ensemble to the leading order.

The next proposition pertains to average relative entropy of Bures-Hall ensemble.
\begin{proposition}\label{prop:BH}
For density matrices $\rho_{\rm{BH}}$ and $\sigma_{\rm{BH}}$ of Bures-Hall ensemble~(\ref{eq:BH}) of dimension $m$ with parameters $n_1=m+\alpha_1$ and $n_2=m+\alpha_2$, respectively, the average relative entropy~(\ref{eq:rem}) is given by
\begin{eqnarray}\label{eq:r2}
\E\!\left[D(\rho_{\rm{BH}}||\sigma_{\rm{BH}})\right]&=&\psi_{0}\!\left(mn_{2}-\frac{m^{2}}{2}\right)-\psi_{0}\!\left(mn_{1}-\frac{m^{2}}{2}+1\right)+\psi_{0}\!\left(n_{1}+\frac{1}{2}\right)\nonumber\\
&&+~\!\frac{n_{2}}{m}\psi_{0}\!\left(n_{2}\right)-\frac{2n_{2}-m}{m}\left(\psi_{0}\!\left(n_{2}-\frac{m}{2}\right)+\psi_{0}\!\left(n_{2}-\frac{m}{2}+\frac{1}{2}\right)\right)\nonumber\\
&&+~\!\frac{\alpha_{2}}{m}\left(\psi_{0}\!\left(\alpha_{2}\right)+2\psi_{0}\!\left(\alpha_{2}+\frac{1}{2}\right)\right)+1.
\end{eqnarray}
\end{proposition}

The limiting formula of Proposition~\ref{prop:BH} in the regime~(\ref{eq:lim}) is similarly obtained by using the asymptotic expansion~(\ref{eq:dga}) as
\begin{eqnarray}\label{eq:BHlim}
\E\!\left[D(\rho_{\rm{BH}}||\sigma_{\rm{BH}})\right]&=&\ln c_{1}+c_{2}\ln c_{2}-\ln\left(c_{1}-\frac{1}{2}\right)-\left(4c_{2}-3\right)\ln\left(c_{2}-\frac{1}{2}\right)\nonumber\\
&&+~\!3\left(c_{2}-1\right)\ln\left(c_{2}-1\right)+1+\mathcal{O}\left(\frac{1}{m}\right).
\end{eqnarray}

Besides states from the same ensemble, relative entropy $D(\rho||\sigma)$ also quantifies the lack of information between $\sigma$ of one ensemble and $\rho$ of a generalized yet distinct ensemble. The proposition below quantifies the average of $D(\rho_{\rm{BH}}||\sigma_{\rm{HS}})$.
\begin{proposition}\label{prop:BH2HS}
For density matrices $\rho_{\rm{BH}}$ of Bures-Hall ensemble~(\ref{eq:BH}) and $\sigma_{\rm{HS}}$ of Hilbert-Schmidt ensemble~(\ref{eq:HS}) of dimension $m$ with parameters $n_1=m+\alpha_1$ and $n_2=m+\alpha_2$, respectively, the average relative entropy~(\ref{eq:rem}) is given by
\begin{eqnarray}\label{eq:r3}
\E\!\left[D(\rho_{\rm{BH}}||\sigma_{\rm{HS}})\right]&=&\psi_{0}\left(mn_{2}\right)-\psi_{0}\left(mn_{1}-\frac{m^{2}}{2}+1\right)+\psi_{0}\left(n_{1}+\frac{1}{2}\right)\nonumber\\
&&-~\!\frac{1}{m}\left(n_{2}\psi_{0}\left(n_{2}\right)-\alpha_{2}\psi_{0}\left(\alpha_{2}\right)\right)+1.
\end{eqnarray}
\end{proposition}

Note that an exact formula of $\E\!\left[D(\rho_{\rm{HS}}||\sigma_{\rm{BH}})\right]$ can be similarly derived, which, despite being less relevant from an information-theoretic perspective, is presented in~(\ref{eq:r4}) for completeness. We also note that a direct use of~(\ref{eq:dga}) leads to the asymptotic formula of Proposition~\ref{prop:BH2HS} in the regime~(\ref{eq:lim}) as
\begin{eqnarray}\label{eq:HS2BHlim}
\E\!\left[D(\rho_{\rm{BH}}||\sigma_{\rm{HS}})\right]&=&\ln c_{1}-\left(c_{2}-1\right)\ln c_{2}-\ln\left(c_{1}-\frac{1}{2}\right)\nonumber\\
&&+\left(c_{2}-1\right)\ln\left(c_{2}-1\right)+1+\mathcal{O}\left(\frac{1}{m}\right).
\end{eqnarray}

\begin{figure}[!h]
\begin{center}
\vspace{-6.4cm}
\hspace{-2.9cm}\includegraphics[width=1.2\linewidth]{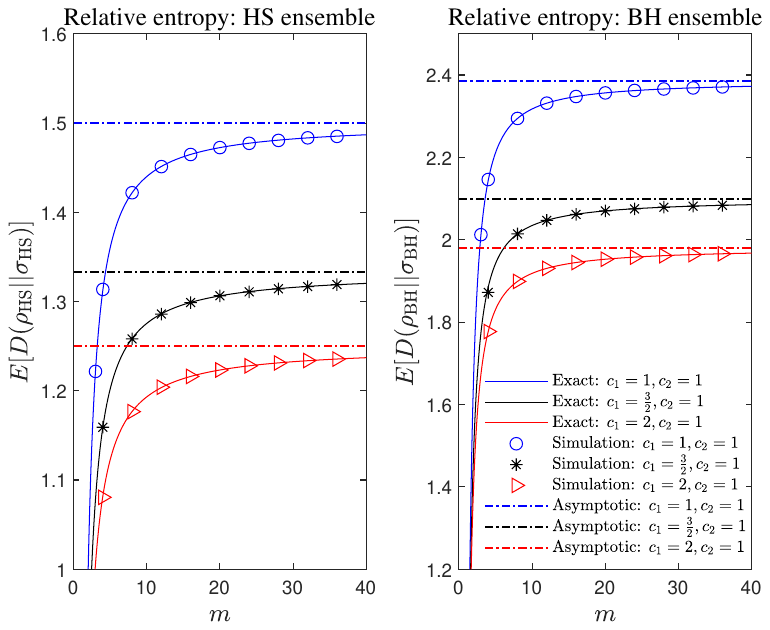}\hspace{-2.9cm}
\vspace{-5.7cm}
\caption{Average relative entropy of states from the same ensemble: analytical results versus simulations. Solid lines are drawn by the exact formulas~(\ref{eq:r1}) and~(\ref{eq:r2}), while dash-dot horizontal lines are drawn by the corresponding limiting expressions~(\ref{eq:HSlim}) and~(\ref{eq:BHlim}). The symbols of circle, asterisk, and triangle, represent numerical simulations.}
\label{fig:1}
\end{center}
\vspace{0cm}
\end{figure}

\begin{figure}[!h]
\begin{center}
\vspace{-4.9cm}
\hspace{-2.9cm}\includegraphics[width=1.2\linewidth]{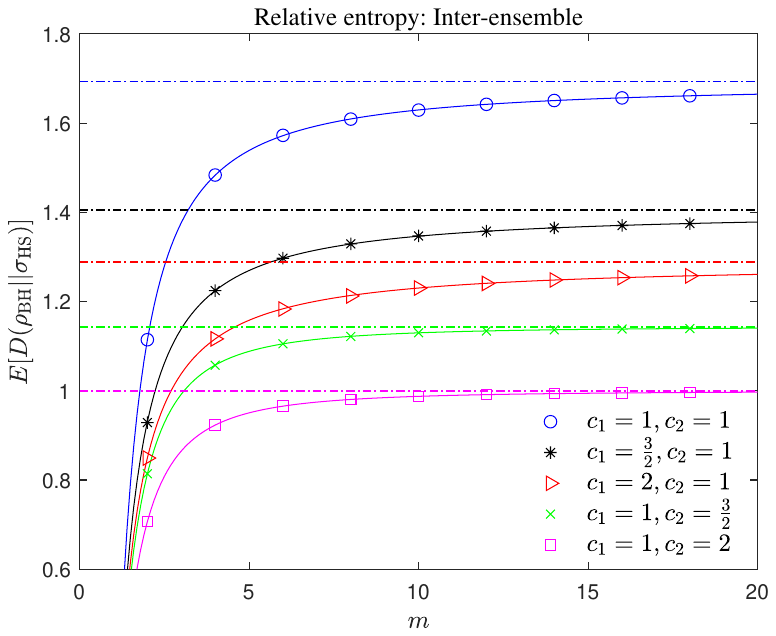}\hspace{-2.9cm}
\vspace{-5.7cm}
\caption{Average relative entropy of states from two different ensembles: analytical results versus simulations. Solid lines are drawn by the exact formula~(\ref{eq:r3}), while dash-dot horizontal lines are drawn by the corresponding limiting expression~(\ref{eq:HS2BHlim}). The symbols of circle, asterisk, triangle, cross, and square represent numerical simulations.}
\label{fig:2}
\end{center}
\vspace{0cm}
\end{figure}

Before presenting proofs to the main results in Section~\ref{sec:proof}, we perform a number of numerical studies as summarized in the two figures. In Figure~\ref{fig:1}, we plot average relative entropy of states from the same ensemble, where the left-hand side and right-hand side subfigures correspond to the Hilbert-Schmidt and Bures-Hall ensembles, respectively. For each ensemble, we consider three different values of parameter $c_{1}$ while fixing $c_{2}$. The numerical simulations, illustrated by the scatters with each averaged over $10^{6}$ realizations of density matrices, match well with exact formulas of Proposition~\ref{prop:HS} and Proposition~\ref{prop:BH}. As system dimension $m$ increases, we observe fast convergence to the dash-dot horizontal lines representing limiting values computed from~(\ref{eq:HSlim}) and~(\ref{eq:BHlim}). It is seen from the figure that for a fixed $c_{2}$ the average relative entropy decreases as $c_{1}$ increases, which agrees with the observation in~\cite{Kudler21}. This phenomenon is consistent with the general principle that relative entropy becomes larger when the two density matrices $\rho$ and $\sigma$ are more distinct~\cite{Chen25}. Intuitively, as $c_{1}$ (or $c_{2}$) increases, the random density matrix $\rho$ (or $\sigma$) approaches a deterministic matrix leading to diminished distinguishability between the two density matrices. The case $c_1=c_2=1$ reaches the largest value of average relative entropy for any given $m$, which corresponds to maximum randomness in the density matrices. In fact, as $c_1$ or $c_2$ increases the monotonic decreasing phenomenon can be analytically verified from the limiting formulas~(\ref{eq:HSlim}) and~(\ref{eq:BHlim}). We also observe that the average relative entropy of Bures-Hall ensemble attains a larger value than the corresponding Hilbert-Schmidt ensemble, which is due to the wider spectral width of Bures-Hall ensemble~\cite{Wei23}.

In Figure~\ref{fig:2}, we plot average relative entropy of states from two different ensembles pertaining to the case in Proposition~\ref{prop:BH2HS}. We consider several values of parameters $c_{1}$ and $c_{2}$. Similarly to Figure~\ref{fig:1}, one observes even faster convergence to the dash-dot horizontal lines that represent limiting constants calculated from~(\ref{eq:HS2BHlim}). As expected, it is seen that for a fixed parameter the average relative entropy decreases as the other parameter increases, where the case $c_1=c_2=1$ corresponds to the largest value of average relative entropy for a given $m$. It is also observed that for $c_{1}$ greater than $c_{2}$ the relative average entropy is greater than that of the case when $c_{1}$ and $c_{2}$ are exchanged. This may indicate improved robustness to parameter changes of Bures-Hall ensemble in distinguishing quantum states than Hilbert-Schmidt ensemble.

\section{Proofs of Main Results}\label{sec:proof}
In this section, we present proofs to the main results in Propositions~\ref{prop:HS}--\ref{prop:BH2HS} on average relative entropy. In Section~\ref{sec:ensemle}, we introduce random state models of Hilbert-Schmidt and Bures-Hall ensembles, and discuss their significance in quantum information processing. In Section~\ref{sec:3prop}, we first evaluate the required average over unitary group of the problem that leads to a factorization of averages of two density matrices before computing the factorized averages in scenarios considered in the three propositions.

\subsection{Generic state ensembles}\label{sec:ensemle}
We outline the density matrix formalism~\cite{Bengtsson}, introduced by von Neumann, from which different generic state ensembles are constructed. Consider a composite system of two subsystems $A$ and $B$ of Hilbert space dimensions $m$ and $n$, respectively. Without loss of generality, one assumes $m\leq n$. A generic state of the bipartite system is written as a linear combination of the random coefficients $c_{i,j}$ and bases of the two subsystems as
\begin{equation}\label{eq:state}
\Ket{\psi}=\sum_{i=1}^{m}\sum_{j=1}^{n}c_{i,j}\Ket{i_{\rm{A}}}\otimes\Ket{j_{\rm{B}}},
\end{equation}
where the coefficients $c_{i,j}$ are independently and identically distributed standard complex Gaussian random variables.

The Hilbert-Schmidt ensemble is the probability density function of the $m\times m$ reduced density matrix $\rho_{\rm{A}}=\tr_{\rm{B}}(\rho)$ of the smaller subsystem $A$, obtained through partial trace of the full density matrix $\rho=\ket{\psi}\bra{\psi}$, as~\cite{Page93,Bengtsson}
\begin{equation}\label{eq:HS}
f_{\rm{HS}}\!\left(\rho_{\rm{A}}\right)=\frac{1}{C_{\rm{HS}}}
\delta\left(1-\tr\rho_{\rm{A}}\right)\det\!\!\!~^{\alpha}(\rho_{\rm{A}})\dd\rho_{\rm{A}},
\end{equation}
where
\begin{equation}\label{eq:cHS}
C_{\rm{HS}}=\frac{\pi^{\frac{1}{2}m(m-1)}}{\Gamma\left(m(m+\alpha)\right)}\prod_{i=1}^{m}\Gamma(i+\alpha)
\end{equation}
and
\begin{equation}\label{eq:a}
\alpha=n-m\geq0
\end{equation}
denotes the dimension difference of the two subsystems. The Hilbert-Schmidt ensemble~(\ref{eq:HS}) is also referred to as fixed-trace Wishart ensemble~\cite{Page93,Ruiz95,HW25} constructed from a Wishart matrix $YY^{\dag}$ as
\begin{equation}\label{eq:maHS}
\rho_{\rm{A}}=\frac{YY^{\dag}}{\tr\left(YY^{\dag}\right)}
\end{equation}
with $Y$ denoting an $m\times n$ matrix of independent complex Gaussian entries.

The Bures-Hall ensemble generalizes the Hilbert-Schmidt ensemble in that its state
\begin{equation}\label{eq:BHs}
\Ket{\varphi}=\Ket{\psi}+\left(U\otimes I_{n}\right)\Ket{\psi}
\end{equation}
is a superposition of the state~(\ref{eq:state}) with a rotation by an $m\times m$ random unitary matrix $U$, where $I_{n}$ denotes an identity matrix of dimension $n$. The probability density function of the unitary matrix $U$ is proportional to $\det\left(I_{m}+U\right)^{2\alpha}$. Taking the partial trace $\rho_{\rm{A}}=\tr_{\rm{B}}(\rho)$ over density matrix $\rho=\Ket{\varphi}\Bra{\varphi}$ of the state~(\ref{eq:BHs}), the Bures-Hall ensemble is written as~\cite{Kumar19,Bengtsson}
\begin{equation}\label{eq:BH}
f_{\rm{BH}}\!\left(\rho_{\rm{A}}\right)=\frac{1}{C_{\rm{BH}}}
\delta\left(1-\tr\rho_{\rm{A}}\right)\det\!\!\!~^{\alpha}(\rho_{\rm{A}})\int_{Z}\ee^{-\tr\left(\rho_{\rm{A}}Z^{2}\right)}\dd Z\dd\rho_{\rm{A}},
\end{equation}
where
\begin{equation}\label{eq:cBH}
C_{\rm{BH}}=\frac{\pi^{m^2}m!~\!2^{-m(m+2\alpha-1)}}{\Gamma\left(m(m+2\alpha)/2\right)}\prod_{i=1}^{m}\frac{\Gamma(i+2\alpha)}{\Gamma(i+\alpha)}
\end{equation}
and $Z$ is an $m\times m$ Hermitian matrix. The matrix model corresponding to the state~(\ref{eq:BHs}) of Bures-Hall ensemble is~\cite{Kumar19}
\begin{equation}\label{eq:maBH}
\rho_{\rm{A}}=\frac{\left(I_{m}+U\right)YY^{\dag}\left(I_{m}+U^{\dag}\right)}{\tr\left(\left(I_{m}+U\right)YY^{\dag}\left(I_{m}+U^{\dag}\right)\right)},
\end{equation}
where $YY^{\dag}$ is a Wishart matrix as in~(\ref{eq:maHS}).

The Dirac delta function $\delta$ in the ensembles~(\ref{eq:HS}) and~(\ref{eq:BH}) reflects the defining property of density matrices~(\ref{eq:dm}) that
\begin{equation}\label{eq:t1}
\tr\rho_{\rm{A}}=1.
\end{equation}
It is important to point out that the two unitarily invariant densities~(\ref{eq:HS}) and~(\ref{eq:BH}) are valid for a non-negative real $\alpha$, and so are the results obtained involving $\alpha$. Note also that we only display matrix-variate densities of the two ensembles that are sufficient for the purpose of this work. For computation of entanglement entropy involving eigenvalue densities of the two ensembles, we refer to~\cite{HWC21,HW25,Page93,Ruiz95,Kumar19,VPO16,Wei17,Wei20,Wei20BHA,Wei20BH,WHW25}.

In principle, other generic state ensembles in the space of density matrices~\cite{Bengtsson} can be constructed such as the fermionic Gaussian ensemble~\cite{Bianchi21,HW23}. Main reasons that the considered ensembles~(\ref{eq:HS}) and~(\ref{eq:BH}) stand out as major ones are discussed below.
\begin{itemize}
\item \textbf{The Hilbert-Schmidt ensemble} corresponds to the simplest model of generic quantum states, where no prior information of the states needs to be assumed. The randomness of the states comes from the assumption of Gaussian distributed coefficients, which correspond to the most non-informative distribution. The ensemble can be thought of as the baseline Gaussian model universal in statistical modelling of an unknown variable. In the investigation of quantum information processing tasks, it is desirable to make use of Gaussian generic states to benchmark the performance.

\item \textbf{The Bures-Hall ensemble} is an improved variant of the Hilbert-Schmidt ensemble that satisfies a few additional properties. The Bures metric, that induces the Bures-Hall ensemble, is the only monotone metric that is simultaneously Fisher adjusted and Fubini-Study adjusted. The Bures metric, related to quantum distinguishability, is the minimal monotone metric. The ensemble is often used as a prior distribution known as Bures prior in reconstructing quantum states from measurements. Generic states from the Hilbert-Schmidt and Bures-Hall ensembles are physical in that they can be generated in polynomial time.
\end{itemize}

\subsection{Computation of average relative entropy}\label{sec:3prop}
We consider eigenvalue decompositions of the density matrices
\begin{equation}
\rho=V\Lambda_{\rho}V^{\dag},~~~~~~~~\sigma=W\Lambda_{\sigma}W^{\dag},
\end{equation}
where the diagonal matrices $\Lambda_{\rho}$, $\Lambda_{\sigma}$ consist of eigenvalues and $V$, $W$ are unitary matrices of eigenvectors. It is seen that the first term of the relative entropy~(\ref{eq:red}),
\begin{equation}
\tr(\rho\ln\rho)=\tr\left(\Lambda_{\rho}\ln\Lambda_{\rho}\right)
\end{equation}
does not depend on the eigenvectors, whereas the second term
\begin{equation}\label{eq:rem2a}
\tr(\rho\ln\sigma)=\tr\left(\Lambda_{\rho}U\ln\Lambda_{\sigma}U^{\dag}\right)
\end{equation}
does depend on eigenvectors through the unitary matrix $U=V^{\dag}W$. Since the densities~(\ref{eq:HS}) and~(\ref{eq:BH}) are unitarily invariant, evaluating the average~(\ref{eq:rem2a}) over eigenvalues and eigenvectors can be performed separately. The average over the latter is shown below via the connection to zonal polynomials~\cite{Khatri66,Forrester,Mathai}.

For an $m\times m$ Hermitian matrix $X$, the $l$-th positive integer power of the trace can be uniquely decomposed as~\cite{Mathai}
\begin{equation}\label{eq:trc}
\tr^{l}\!\left(X\right)=\sum_{\kappa}C_{\kappa}(X),
\end{equation}
where the sum of zonal polynomials $C_{\kappa}(X)$ is over partitions $\kappa=(\kappa_{1},\kappa_{2},\dots,\kappa_{m})$ of $l$ into no more than $m$ parts
\begin{equation}
\kappa_{1}+\kappa_{2}+\dots+\kappa_{m}=l,~~~~~~\kappa_{1}\geq\kappa_{2}\geq\dots\geq\kappa_{m}\geq0.
\end{equation}
The zonal polynomial $C_{\kappa}(X)$ is a symmetric polynomial of degree $l$ in the $m$ eigenvalues $\{x_i\}_{i=1}^{m}$ of $X$ given by
\begin{equation}\label{eq:czp}
C_{\kappa}(X)=\chi_{\kappa}(1)\chi_{\kappa}(X),
\end{equation}
where
\begin{equation}\label{eq:czp1}
\chi_{\kappa}(1)=\frac{l!\prod_{1\leq i<j\leq m}(\kappa_{i}-\kappa_{j}-i+j)}{\prod_{j=1}^{m}(\kappa_{j}+m-j)!}
\end{equation}
is the dimension of the representation of the symmetric group and
\begin{equation}\label{eq:czp2}
\chi_{\kappa}(X)=\frac{\det\left(x_{i}^{\kappa_{j}+m-j}\right)}{\det\left(x_{i}^{m-j}\right)}
\end{equation}
is character of the representation. In fact, the character~(\ref{eq:czp2}) is the Schur polynomial, which can be written as a determinant of elementary symmetric polynomials
\begin{equation}
e_{k}\left(x_{1},\dots,x_{m}\right)=\sum_{1\leq i_1<\cdots<i_k\leq m}x_{i_1}x_{i_2}\cdots x_{i_k}
\end{equation}
as~\cite{Macdonald}
\begin{equation}\label{eq:czp3}
\chi_{\kappa}(X)=\det\left(e_{\kappa'-i+j}\left(x_{1},x_{2},\dots,x_{m}\right)\right)_{i,j=1}^{l(\kappa')}
\end{equation}
with $l(\kappa)$ and $\kappa'$ respectively denoting the length and the conjugate of a partition $\kappa$.

Of interest to this work is the following integral identity of zonal polynomials involving two $m\times m$ Hermitian matrices $X$, $Y$ over the invariant measure $\dd U$ of the unitary group $U(m)$ as~\cite{Khatri66,Forrester}
\begin{equation}\label{eq:AUBU}
\int_{U(m)}C_{\kappa}\!\left(XUYU^{\dag}\right)\dd U=\frac{C_{\kappa}\!\left(X\right)C_{\kappa}\!\left(Y\right)}{C_{\kappa}\!\left(I_{m}\right)}
\end{equation}
valid for any partition $\kappa$. As a result, the unitary matrix in~(\ref{eq:rem2a}) is integrated out as
\begin{eqnarray}
\int_{U(m)}\tr(\rho\ln\sigma)\dd U &=& \int_{U(m)}C_{1}\!\left(\Lambda_{\rho}U\ln\Lambda_{\sigma}U^{\dag}\right)\dd U \label{eq:U1}\\
&=&\frac{C_{1}\!\left(\Lambda_{\rho}\right)C_{1}\!\left(\ln\Lambda_{\sigma}\right)}{C_{1}\!\left(I_{m}\right)} \label{eq:U2} \\
&=&\frac{1}{m}\tr\rho\tr\left(\ln\sigma\right) \label{eq:U3} \\
&=&\frac{1}{m}\tr\left(\ln\sigma\right). \label{eq:U4}
\end{eqnarray}
In obtaining~(\ref{eq:U1}), we have employed the result~(\ref{eq:trc}) for $l=1$,
\begin{equation}
\tr(X)=C_{1}(X),
\end{equation}
where the only partition is
\begin{equation}\label{eq:k1}
\kappa_{1}=1,~~~~\kappa_{2}=\dots=\kappa_{m}=0.
\end{equation}
The result~(\ref{eq:U2}) is an application of~(\ref{eq:AUBU}). The equality~(\ref{eq:U3}) follows from the identity
\begin{equation}
C_{1}\!\left(I_{m}\right)=\chi_{\kappa}(1)\chi_{\kappa}\!\left(I_{m}\right)=m,
\end{equation}
which is established by inserting the partition~(\ref{eq:k1}) into~(\ref{eq:czp1}) and~(\ref{eq:czp3}) leading respectively to
\begin{equation}
\chi_{\kappa}(1)=1,~~~~~~\chi_{\kappa}\!\left(I_{m}\right)=e_{1}(1,1,\dots,1)=m.
\end{equation}
Finally, (\ref{eq:U4}) is due to the fact that $\rho$ is a density matrix~(\ref{eq:t1}).

The above machinery of evaluating unitary integrals~(\ref{eq:AUBU}) through the connection to zonal polynomials~(\ref{eq:trc}) will be a key ingredient in computing higher-order moments of relative entropy. However, if the only goal is to compute the average relative entropy, the result~(\ref{eq:U4}) can be directly obtained by a standard fact of Weingarten calculus of a unitary matrix $U=\left(u_{ij}\right)_{i,j=1}^{m}$,
\begin{equation}
\int_{U(m)}u_{ij}u_{kl}^{\dag}\dd U=\frac{1}{m}\delta_{ik}\delta_{jl}
\end{equation}
as
\begin{eqnarray}
\int_{U(m)}\tr(\rho\ln\sigma)\dd U &=& \int_{U(m)}\tr\left(\Lambda_{\rho}U\ln\Lambda_{\sigma}U^{\dag}\right)\dd U \\
&=&\sum_{i,j=1}^{m}\rho_{i}\ln\sigma_{j}\int_{U(m)}|u_{ij}|^{2}\dd U \\
&=&\frac{1}{m}\sum_{i=1}^{m}\rho_{i}\sum_{j=1}^{m}\ln\sigma_{j} \\
&=&\frac{1}{m}\tr\left(\ln\sigma\right), \label{eq:U5}
\end{eqnarray}
where $\{\rho_i\}_{i=1}^{m}$ and $\{\sigma_i\}_{i=1}^{m}$ are the set of eigenvalues of $\rho$ and $\sigma$, respectively.

Either~(\ref{eq:U4}) or~(\ref{eq:U5}) now leads to the result
\begin{equation}\label{eq:rem3}
\E\!\left[\tr(\rho\ln\sigma)\right]=\frac{1}{m}\E\!\left[\tr(\ln\sigma)\right],
\end{equation}
which is the starting point of following proofs.

\subsubsection*{Proof of Proposition~\ref{prop:HS}}
For the Hilbert-Schmidt ensemble~(\ref{eq:HS}), by~(\ref{eq:rem}) and~(\ref{eq:rem3}) the remaining task is to compute
\begin{equation}
\E\!\left[\tr(\ln\sigma_{\rm{HS}})\right]
\end{equation}
that amounts to
\begin{eqnarray}
\E\!\left[\tr(\ln\sigma_{\rm{HS}})\right] &=& \E\!\left[\ln\left(\det\sigma_{\rm{HS}}\right)\right] \\
&=&\frac{1}{C_{\rm{HS}}}\int\delta\left(1-\tr\sigma_{\rm{HS}}\right)\frac{\dd}{\dd\alpha}\det\!\!\!~^{\alpha}(\sigma_{\rm{HS}})\dd\sigma_{\rm{HS}} \\
&=&\frac{1}{C_{\rm{HS}}}\frac{\dd}{\dd\alpha}C_{\rm{HS}}, \label{eq:dHSc}
\end{eqnarray}
where one has made use of the definition
\begin{equation}
C_{\rm{HS}}=\int\delta\left(1-\tr\sigma_{\rm{HS}}\right)\det\!\!\!~^{\alpha}(\sigma_{\rm{HS}})\dd\sigma_{\rm{HS}}.
\end{equation}
The result~(\ref{eq:dHSc}) requires $\alpha$ derivative of the constant~(\ref{eq:cHS}) computed as
\begin{eqnarray}
\frac{\dd}{\dd\alpha}C_{\rm{HS}} &=& \left(-m\psi_{0}(m(m+\alpha))+\sum_{i=1}^{m}\psi_0(i+\alpha)\right)C_{\rm{HS}} \\
&=& \left(-m\psi_{0}(mn)+n\psi_{0}(n)-\alpha\psi_0(\alpha)-m\right)C_{\rm{HS}}, \label{eq:dHScr}
\end{eqnarray}
where we have utilized the definitions~(\ref{eq:dg}),~(\ref{eq:a}), and the summation formula~\cite{Wei17,HW25}
\begin{equation}\label{eq:sum}
\sum_{i=1}^{m}\psi_0(i+\alpha)=(m+\alpha)\psi_{0}(m+\alpha)-\alpha\psi_0(\alpha)-m.
\end{equation}
Inserting~(\ref{eq:dHScr}) into~(\ref{eq:dHSc}), one arrive at
\begin{equation}\label{eq:rem3HS}
\E\!\left[\tr(\ln\sigma_{\rm{HS}})\right]=-m\psi_{0}(mn)+n\psi_{0}(n)-\alpha\psi_0(\alpha)-m.
\end{equation}
Finally, putting together the results~(\ref{eq:rem}),~(\ref{eq:HSm}),~(\ref{eq:rem3}), and~(\ref{eq:rem3HS}) while keeping in mind that the respective parameters in~(\ref{eq:HSm}) and~(\ref{eq:rem3HS}) are $n_{1}$ and $n_{2}$ leads to the average relative entropy formula~(\ref{eq:r1}) in Proposition~\ref{prop:HS}.

\subsubsection*{Proof of Proposition~\ref{prop:BH}}
For the Bures-Hall ensemble~(\ref{eq:BH}), the remaining task is also to compute
\begin{equation}
\E\!\left[\tr(\ln\sigma_{\rm{BH}})\right]
\end{equation}
that, similarly to~(\ref{eq:dHSc}), boils down to computing $\alpha$ derivative of the constant~(\ref{eq:cBH}) as

\begin{eqnarray}
\E\!\left[\tr(\ln\sigma_{\rm{BH}})\right] &=& \E\!\left[\ln\left(\det\sigma_{\rm{BH}}\right)\right] \nonumber \\
&=&\frac{1}{C_{\rm{BH}}}\int\delta\left(1-\tr\sigma_{\rm{BH}}\right)\frac{\dd}{\dd\alpha}\det\!\!\!~^{\alpha}(\sigma_{\rm{BH}})\int_{Z}\ee^{-\tr\left(\sigma_{\rm{BH}}Z^{2}\right)}\dd Z\dd\sigma_{\rm{BH}} \nonumber \\
&=&\frac{1}{C_{\rm{BH}}}\frac{\dd}{\dd\alpha}C_{\rm{BH}} \nonumber \\
&=&-2m\ln2-m\psi_{0}\!\left(\frac{m}{2}(m+2\alpha)\right)+2\sum_{i=1}^{m}\psi_{0}(i+2\alpha)-\sum_{i=1}^{m}\psi_{0}(i+\alpha) \nonumber\\
&=&-m\psi_{0}\!\left(mn-\frac{m^{2}}{2}\right)-n\psi_{0}\!\left(n\right)-\alpha\left(\psi_{0}\!\left(\alpha\right)+2\psi_{0}\!\left(\alpha+\frac{1}{2}\right)\right) \nonumber\\
&&+\left(2n-m\right)\left(\psi_{0}\!\left(n-\frac{m}{2}\right)+\psi_{0}\!\left(n-\frac{m}{2}+\frac{1}{2}\right)\right)-m, \label{eq:rem3BH}
\end{eqnarray}
where we have utilized the definitions~(\ref{eq:dg}),~(\ref{eq:a}), the summation formula~(\ref{eq:sum}), and the identity~\cite{Brychkov}
\begin{equation}
2\psi_{0}(2x)=\psi_{0}\!\left(x+\frac{1}{2}\right)+\psi_{0}(x)+2\ln2.
\end{equation}
Inserting the results~(\ref{eq:BHm}),~(\ref{eq:rem3}), and~(\ref{eq:rem3BH}) into~(\ref{eq:rem}) before replacing parameters $n$ in~(\ref{eq:BHm}) and~(\ref{eq:rem3BH}) respectively by $n_{1}$ and $n_{2}$, we arrive at the formula~(\ref{eq:r2}) in Proposition~\ref{prop:BH}.

\subsubsection*{Proof of Proposition~\ref{prop:BH2HS}}
For average relative entropy between Bures-Hall and Hilbert-Schmidt ensembles, the corresponding formulas can be read off from the obtained results~(\ref{eq:BHm}),~(\ref{eq:rem3}), and~(\ref{eq:rem3HS}) as
\begin{eqnarray}
\E\!\left[D(\rho_{\rm{BH}}||\sigma_{\rm{HS}})\right]&=&\E\!\left[\tr(\rho_{\rm{BH}}\ln\rho_{\rm{BH}})\right]-\E\!\left[\tr(\rho_{\rm{BH}}\ln\sigma_{\rm{HS}})\right]\\
&=&\E\!\left[\tr(\rho_{\rm{BH}}\ln\rho_{\rm{BH}})\right]-\frac{1}{m}\E\!\left[\tr(\ln\sigma_{\rm{HS}})\right]\\
&=&\psi_{0}\!\left(mn_{2}\right)-\psi_{0}\!\left(mn_{1}-\frac{m^{2}}{2}+1\right)+\psi_{0}\!\left(n_{1}+\frac{1}{2}\right)\nonumber\\
&&-~\!\frac{1}{m}\left(n_{2}\psi_{0}\!\left(n_{2}\right)-\alpha_{2}\psi_{0}\!\left(\alpha_{2}\right)\right)+1,
\end{eqnarray}
which is the formula~(\ref{eq:r3}) in Proposition~\ref{prop:BH2HS}.

Similarly, an exact formula of average relative entropy of the case $D(\rho_{\rm{HS}}||\sigma_{\rm{BH}})$ can be extracted as
\begin{eqnarray}\label{eq:r4}
\E\!\left[D(\rho_{\rm{HS}}||\sigma_{\rm{BH}})\right]&=&\psi_{0}\!\left(mn_{2}-\frac{m^{2}}{2}\right)-\psi_{0}\!\left(mn_{1}+1\right)+\psi_{0}\!\left(n_{1}\right)\nonumber\\
&&+~\!\frac{n_{2}}{m}\psi_{0}\!\left(n_{2}\right)-\frac{2n_{2}-m}{m}\left(\psi_{0}\!\left(n_{2}-\frac{m}{2}\right)+\psi_{0}\!\left(n_{2}-\frac{m}{2}+\frac{1}{2}\right)\right)\nonumber\\
&&+~\!\frac{\alpha_{2}}{m}\left(\psi_{0}\!\left(\alpha_{2}\right)+2\psi_{0}\!\left(\alpha_{2}+\frac{1}{2}\right)\right)+\frac{m+2n_{1}+1}{2n_{1}}.
\end{eqnarray}

\backmatter
\bmhead{Acknowledgment} The work of Lu Wei was supported by the U.S. National Science Foundation (2306968) and the U.S. Department of Energy (DE-SC0024631).



\end{document}